\documentclass[twocolumn]{aastex631}

\shorttitle{Binary evolution in second-generation star-formation environments}
\shortauthors{Rozner \& Perets}

\usepackage{graphicx}	
\usepackage{amsmath}	
\usepackage{amssymb}	
\usepackage{color}
\usepackage{float}
\usepackage{ulem}
\usepackage{soul}
\usepackage{lipsum}
\usepackage[T1]{fontenc}

\DeclareRobustCommand{\VAN}[3]{#2}
\let\VANthebibliography\thebibliography
\def\thebibliography{\DeclareRobustCommand{\VAN}[3]{##3}\VANthebibliography}

\usepackage{graphicx}	
\usepackage{amsmath}	
\usepackage{braket}

\begin{document}


\title{Binary evolution, gravitational-wave mergers and explosive transients\\ in multiple-populations gas-enriched globular-clusters}


\email{morozner@campus.technion.ac.il}

\author[0000-0002-2728-0132]{Mor Rozner}
\affiliation{Technion - Israel Institute of Technology, Haifa, 3200002, Israel}


\author[0000-0002-5004-199X]{Hagai B. Perets}
\affiliation{Technion - Israel Institute of Technology, Haifa, 3200002, Israel}

\begin{abstract}
Most globular clusters (GCs) show evidence for multiple stellar populations, suggesting the occurrence of several distinct star-formation episodes. The large fraction of second population (2P) stars observed requires a very large 2P gaseous mass to have accumulated in the cluster core to form these stars. Hence the first population of stars (1P) in the cluster core has had to become embedded in 2P gas, just prior to the formation of later populations. Here we explore the evolution of binaries in ambient 2P gaseous media of multiple-population GCs. We mostly focus on black hole binaries and follow their evolution as they evolve from wide binaries towards short periods through interaction with ambient gas, followed by gravitational-wave (GW) dominated inspiral and merger. We show this novel GW-merger channel could provide a major contribution to the production of GW-sources. We consider various assumptions and initial conditions and calculate the resulting gas-mediated change in the population of binaries and the expected merger rates due to gas-catalyzed  GW-inspirals. For plausible conditions and assumptions, we find an expected GW merger rate observable by aLIGO of the order of up to a few tens of $\rm{Gpc^{-3} yr^{-1}}$, and an overall range for our various models of $0.08-25.51  \ \rm{Gpc^{-3} yr^{-1}}$. Finally, our results suggest that the conditions and binary properties in the early stage of GCs could be critically affected by gas-interactions and may require a major revision in the current modeling of the evolution of GCs. 
\end{abstract}


\section{Introduction} 
Stars are thought to form following the collapse of giant molecular clouds (GMCs), and further grow and evolve through accretion from, and interaction with the GMC ambient gaseous environment during their early evolution, of up to a few Myrs. Following the gas dispersal and depletion, the later long-term evolution of stars and multiple systems is thought to be dominated by their gas-free stellar evolution and their dynamical interactions with other stellar companions and/or stars in the cluster. 
However, some environments can be replenished with gas leading to late epochs of stellar and binary evolution of stars embedded in gas. Already decades ago, \cite{BahcallOstriker1976} have suggested that stellar compact objects can interact with gaseous disks around massive black holes (active galactic nuclei; AGNs), accrete and give rise to X-ray flarings.  \cite{ost83} suggested that stars and compact objects embedded in AGNs disks can accrete gas from the ambient gaseous medium, grow to Chandrasekhar mass and explode as type Ia supernovae (SNe), and later \cite{Artymowicz1993} discussed accretion onto stars in AGN disks giving rise to massive stars exploding as core-collapse (CC) SNe and polluting the AGN disks. 

The dynamical evolution of {\emph binary} gravitating objects embedded in a large-scale gaseous environment could be altered through gas-dynamical friction and accretion that change their orbit and masses and potentially catalyze their merger. We have first discussed binary evolution in gaseous media in the context of catalyzed mergers of binary planetesimals in a protoplanetary disks \citep{PeretsMurrayClay2011,GrishinPerets2016}, and later in the context of compact object binaries in AGN disks \citep{McKernan2012}, where the latter have been extensively studied since then \citep[e.g.][and references therein]{Stone2017,McKernan2018,Roupas2019,Tagawa2020}. \cite{BaruteauCuadraLin2011} explored the evolution of binary {\emph main-sequence stars} (MS) in gas disks around massive black holes (MBHs), suggesting they harden and merger through the interaction with the gas.  Various studies followed the evolution of pre-MS/MS binaries embedded in gas just following their formation during the star-formation epoch of stars in molecular clouds/young clusters, also suggesting that binaries can shrink and merge through the process \citep{gor+96,er+09,Korntreff2012}. It was also suggested that the evolution of embedded binaries could be driven by the formation of a circumbinary disk, which torques the binary. The evolution of binaries in circumbinary disks have been more extensively studied over a wide range of scales from planets, to stars and MBHs (though typically not in the context of a large-scale gaseous environment), but the exact evolution and even the direction of the binary migration in such circumbinary disks are still debated \citep[e.g.][and references therein]{art+91,art+94,bat00,tan+17,Moody2019,mun+19,duf+20,mun+20}.

Although the evolution of stars, binaries and compact objects embedded in gaseous (typically AGN) disks near MBHs have been extensively studied over the last few years, other gas-embedded stellar environments received far less attention. Here and in a companion paper \citep{Perets2022} we study the evolution of single and binary compact-object binaries in the early gas-rich environments that likely existed in multiple-population globular clusters (GCs) and other young massive clusters (YMCs). We also briefly discuss other (non compact-object - main-sequence and evolved) stars and binaries in such environments, but postpone detailed study of the latter to future exploration.

As we discuss below, such gas-rich environments are likely to be far more ubiquitous than AGN disks and potentially play a key role in the the production of compact binaries, binary mergers, gravitational-waves sources and explosive transients.  

For decades, GCs were thought to host simple stellar populations formed through a single star-formation episode. However, detailed observations over the last decade \citep[see e.g.][and references therein]{Carretta2009,Bastian2018} have shown that the vast majority of galactic GCs host multiple stellar populations showing different light elements content. The origins of multiple populations have been extensively studied, but no clear solution has yet been found \citep[see][for summaries of the scenarios and their caveats]{Renzini2015, Bastian2018, Gratton2019}. The current thought  is that GCs experienced two or more star formation episodes, in which second generation/population (2P) stars formed from processed (2P) gas lost from earlier generation/population (1P) stars, and/or accreted external gas. Kinematics show that 2P stars are more centrally concentrated and were likely formed in the inner region of the GC where the 2P gas is expected to have accumulated.

While the source of the 2P gas is debated, the late formation of 2P stars require that tens up to hundreds of Myrs after their formation, 1P stars had become embedded in a highly gas-rich environment that later produced the 2P stars. 
The evolution of stars, binaries and compact objects embedded in gas could therefore be significantly altered in such gaseous environments, following similar processes as discussed for AGN disks and pre-MS stars embedded in the progenitor GMCs. Such processes were little studied in the context gas-embedded multiple-population GCs \citep[but see works by us and others on some aspects of such evolution]{Vesperini2010,Maccarone2012,Leigh2013,Leigh2014,Roupas2019,Perets2022} which is the focus of the the study below. In particular, in this paper, we introduce the effect of gas-catalyzed hardening (shrinkage of the orbit) of binaries in GCs, and discuss its implications for GCs (and YMCs) binary population and binary mergers, the production of GW sources, and the formation of other merger products, compact binaries and explosive transient events catalyzed by binary interactions with gas.

In section \ref{sec:later generation} we briefly discuss the gas replenishment in multiple-population GCs. In section \ref{sec:hardening processes} we describe the hardening processes of binaries in globular clusters due to gas-dynamical friction, and its relation to dynamical hardening by stars and GW inspirals. In section \ref{sec: results} we introduce our results: in subsection \ref{subsec: separation evolution} we focus on the evolution of an individual binary under the effect of gas hardening
and in subsection \ref{subsec: mergers rate} we estimate the expected merger rate from the channel we proposed. In section \ref{subsec: discussion} we discuss our results and additional implications. In section \ref{subsec: summary} we summarize and conclude. 

\section{Multiple stellar populations and early gas-replenishment in GCs}
\label{sec:later generation}
As discussed above and in \cite{Perets2022}, gas could be replenished in GCs (and YMCs) through mass lost from evolved stars and binaries and/or through accretion of external gas onto the clusters (see a detailed review in \citealp{Bastian2018}). 

 The formation channel sets the amount of gas and hence the dynamics and evolution of embedded stars/binaries. Given the correlation between the fractions of 2P stars and GCs properties, it is likely that a large fraction of 2P stars correspond to higher masses of the clusters, larger escape velocities \citep{MastrobuonoBattisiPerets2020}, and hence larger mass of replenished gas.

Given the observed kinematics and concentrations of 2P stars, and theoretical models for the formation and evolution of 2P stars, it is thought that the replenished gas is concentrated in the central part of GCs, where 2P stars are concentrated.  It is likely that the remnant angular momentum of replenished gas gives rise to the formation of 2P  in gaseous disks, rather than spherical distribution \citep{Bekki2010,Bekki2011,MastrobuonoBattistiPerets2013,MastrobuonoBattisiPerets2016}.

The total mass of 2P gas in GCs is highly uncertain, but given reasonable assumptions on the relation between the gas and the observed populations of 2P stars in GCs one can provide an estimate the amount of replenished gas and its density. 
 The typical gas density in star forming regions is usually constrained in the range 
$10^2-10^6 \ M_\odot \rm pc^{-3}$ \citep{Leigh2014}. Estimates for the 2P gas densities could be obtained from a simple order of magnitude calculations, assuming 2P stars were formed from replenished gas.
The gas density is then $\rho_g\sim M_{g}/V_{\rm 2P}$ where $M_g$ is the mass of the gas and $V_{\rm 2P}$ is the typical volume in which the 2P stars reside. 
Following \cite{Bekki2017}, $M_{\rm 2P}\sim 10^5 M_\odot$ and $\epsilon_g=0.3$, then $M_g\sim 3\times 10^5 M_\odot$, where $\epsilon$ is the star-formation efficiency. 
 The infalling replenished gas is likely concentrated in a compact region in the  central parts of GCs, such that the typical effective radius that encloses the 2P population is of the order of $1 \ \rm{pc}$ \citep{Bekki2017}. Taken together, the typical density of the replenished gas is $\sim 3\times 10^5 \ M_\odot \ \rm{pc}^{-3}$, which lies within the expected range for gas densities in star-forming regions. 
 From this density, we will consider scaling to different gas masses, considering $R_{\rm core}=1 \ \rm{pc}$ and take $\rho_{g}\sim M_{\rm g}/R_{\rm core}^3$ accordingly. In particular, as we discuss below, the 2P gas is likely enclosed in a disk-like configuration, in which case the expected gas densities are higher. 
 A priori, the binary hardening releases energy that could heat the gas significantly, but from a crude calculation, the cooling rate is high enough to compensate for it (see also \citealp{Tagawa2020} for a similar calculation in AGN disks). We also note that the possible production of jets could potentially unbind gas from the disk \citep{Soker2016,Tagawajet}, but the study of this possibility is beyond the scope of the current paper. 

The total amount of gas is depleted in time, due to formation of stars and/or accretion onto stars, and later gas ejection through possible radiation pressure processes and SNe. For simplicity we assume an exponential decay, i.e. $\rho_g(t)=\rho_{g,0}\exp(-t/\tau_{\rm gas})$ and consider several possible options for the gas lifetime, to account for uncertainties in the possible gas-depletion processes involved.

\subsection{Disk configuration}\label{subsec: disk configuration}

Gas replenishment leading to the formation of 2P stars in GCs might form a disk-like structure in the cluster nuclei (e.g. \citealp{Bekki2010,MastrobuonoBattistiPerets2013}). 

Following \cite{Bekki2010}, we consider a flat disk, i.e. with a constant aspect ratio. We estimate the aspect ratio by $h/r\sim c_s/v_K$ where $v_K=\sqrt{G(M_{\rm gas}+M_{\star})/R_{\rm core}}$ is the typical velocity in the central parsec. The speed of sound $c_s=\sqrt{k_BT_{\rm gas}/\mu m_p}$ ranges between $0.1-10 \ \rm{km/sec}$ (e.g. \citealp{Bekki2010,Leigh2013}), in correspondence to the gas temperature $T_{\rm gas}$, such that $c_s \approx 0.6 \ \rm{km/sec}$ corresponds to temperature of $100 \ K$, which is the typical temperature in star formation areas, where $\mu=2.3$, 
and $m_p$ is the proton mass.
Exponential disk models were also considered \citep{HenaultBrunet2015}, but here we focus on simple models.
 
In our fiducial model, we consider $c_s=10 \ \rm{km/sec}$, unless stated otherwise. 
Then, the aspect ratio $h/r\approx 0.23$. Following \cite{Bekki2010}, we consider a
velocity dispersion of $\sigma_{\rm disk}=10 \ \rm{km/sec}$ for stars embedded in the disk.
As a conservative assumption, we consider the stellar/massive objects density in the disk to be the same as in the core i.e. $n_{\star, disk}\approx n_\star=10^5 \ \rm{pc^{-3}}$. However, it should be noted that due to gas dynamical friction, stars will migrate and experience inclination damping, and the effective density in the disk is expected to be higher (e.g. \citealp{Artymowicz1993,Leigh2014,GrishinPerets2016}).

We can estimate the volume ratio between the disk and the core volume by $\pi R_{\rm core}^2 h/(4\pi R_{\rm core}^3/3)\sim 0.75h/r$. Then, under the assumption that all the second generation gas is concentrated in the disk, we get a typical gas density of $\rho_{\rm g,disk}\sim 1.74\times 10^6 \ M_\odot \ \rm{pc^{-3}}$.  
The fraction of stars in the disk will change for thinner/thicker disks correspondingly.

The evolution of binaries in disks differs in several aspects from the evolution in a spherical configuration. For our discussion, the major ones are: the velocity dispersion decreases, the gas density increases and the total number of stars contained in the disk is only the volumetric fraction of the disk compared with the volume of the spherical core. The fraction might change with time due to the interaction with gas.

\section{Dynamics of binaries and their interaction with gas: binary hardening and mergers}\label{sec:hardening processes}

Binaries embedded in gas interact with it, exchange angular momentum and energy and possibly accrete gas. These processes are quite complex; here we focus on the interaction through gas-dynamical friction (GDF), while other suggested processes for interaction with gas are discussed in subsec. \ref{subsec: other models}. 

Besides interaction with gas, binaries in GCs can interact with other stars through dissipative effects such as GW inspirals or tidal evolution and through dynamical encounters with other stars through three (or more)-body encounters \citep{Heggie1975}.

The semimajor axis (SMA) of a given massive binary in a gas-enriched environment evolves through the combined effect of the  above-mentioned processes:

\begin{align}
\frac{da_{\rm bin}}{dt}= \frac{da_{\rm bin}}{dt}\bigg|_{\rm 3-body}+\frac{da_{\rm bin}}{dt}\bigg|_{\rm GDF}+\frac{da_{\rm bin}}{dt}\bigg|_{\rm GW}
\end{align}
where $a_{\rm bin}$ is the binary SMA.

A priori, all the three mechanisms contribute to the evolution of the SMA. However, in practice, each of these process dominates in a specific regime, and can be typically neglected in other regimes. Binaries could shrink to shorter periods (harden) due to the effect of gas-interaction or GWs inspiral, and get harder or softer (wider) due to three-body interactions with other GC stars. As we discuss in the following, the evolution of hard binaries is dominated by gas-interactions at large separations and by GW-emission at small separation, while dynamical hardening and softening through three-body encounters \citep{Heggie1975} can be neglected in these regimes. Nevertheless, binary softening and evaporation before the gas-replenishment episode can destroy the widest binaries in the clusters, and hence determine the largest possible initial SMAs for binaries in the cluster at the beginning of the gas-interaction epoch. Moreover, it could play a role in hardening binaries that did not merge within the gas epoch.

The interaction with gas can also give rise to the formation of new wide binaries through two-body and three-body encounters in gas \citep{GoldreichSari2002,Tagawa2020}, allowing for replenishment of binaries in clusters.

In the following we discuss these various processes, while we neglect the effect of direct accretion onto compact objects and their growth, which is beyond the scope of the current paper (though generally such accretion, if effective likely further accelerates binary hardening \citep[e.g.][]{Roupas2019}.

\subsection{Hardening and softening through dynamical encounters with stars}
Due to interactions with other stars, hard binaries tend to get harder, while soft binaries tend to get softer \citep{Heggie1975}; see updated discussion and overview of these issues in \cite{gin21a,gin21b}.
Hence, in the absence of a gaseous environment stellar dynamical hardening  plays an important role in binary evolution and in catalyzing binary mergers.

\subsubsection{Hard binaries} 
For hard binaries, the dynamical hardening rate (up to order unity corrections calibrated usually from numerical simulations) is given by \citep{Spitzer1987}

\begin{align}\label{eq:stellar hardening}
\frac{da_{\rm bin}}{dt}\bigg|_{3-body}=- \frac{2\pi Gn_\star m_{\rm pert}(2m+m_{\rm pert})a_{\rm bin}^2}{mv_\infty}
\end{align}

\noindent 
where we consider a binary with equal mass components, $m=m_1=m_2$ and an external perturber with mass $m_{\rm pert}$. For interactions with other massive objects only, $n_\star$ and $m_{\rm pert}$ should be taken as $n_\bullet$ and $\bar m_\bullet$ correspondingly.

\subsubsection{Soft binaries}\label{subsec: soft binaries}

A binary is called a soft binary if its energy is lower than $\bar m \sigma^2$. This condition sets a critical SMA

\small
\begin{align}
&a_{\rm SH}= \frac{2Gm^2}{\bar m \sigma^2}\approx \\ \nonumber
&\approx 
200.53 \rm{AU} \left(\frac{m}{10 \ M_\odot}\right)^2\left(\frac{43.2 \ \rm{km/sec}}{\sigma}\right)^2\left(\frac{0.5\ M_\odot}{\bar m}\right)
\end{align}

\normalsize
As can be seen, massive stars tend to be hard relative to the background stars in the cluster, due to the scaling $a_{\rm SH}\propto m^2/\bar m$. Hence, one should define the hardness of massive binaries relative to both low mass and high mass stars, in particular, the latter will give rise to softer binaries. We then get the following modified expression \citep{Quinlan1996,KritosCholis2020},

\begin{align}
a_{\rm SH, \bullet}\approx \frac{Gm}{4\sigma^2}\approx 1.25 \rm{AU} \left(\frac{m}{10 \ M_\odot}\right)\left(\frac{43.2 \ \rm{km/sec}}{\sigma}\right)^2
\end{align}

Soft wide binaries are prone to destruction due to encounters with other stars. The dynamical evolution of massive binaries is dominated by interactions with other massive stars and their number density in the core is elevated due to mass segregation \citep{SigurdssonPhinney1995}. 

As to bracket the effect of softening, we consider two possibilities. (1) Softening is dominated by encounters with stellar BHs, where we assume the number density of such objects to be $n_{\rm \rm b}=n_\bullet=10^3 \ \rm{pc^{-3}}$, due to mass segregation to the core, where $\bar m_\bullet=10 \ M_\odot$ (see a discussion in \citealp{MillerHamilton2002}). (2) Softening is dominated by low-mass $0.5 \ M_\odot$ mass stars, if the cluster is not well segregated, and the $n_{\rm b}=n_\star=10^5 \ \rm{pc^{-3}}$.

Hence, the typical lifetime of a soft massive binary is given by 
(e.g. \citealp{BinneyTremaine2008}), 

\begin{align}
\tau_{\rm evap,massive}&\approx  \frac{(m_1+m_2)\sigma}{16\sqrt{\pi}n_{\rm b}\bar {m}^2_b G a \ln \Lambda}
\end{align}

\noindent
where $\ln \Lambda$ is the Coulomb logarithm and $n_b$ and $\bar m_b$ are the number density and the mass of the background stars and change according to our choice between (1) and (2). 
The widest binaries that survive evaporation until the formation of second generation stars, signed as $\tau_{\rm SG}$, taken here to be $100 \ \rm{Myr}$

\begin{align}\label{eq:widest}
a_{\rm widest}&=\max\left\{a_{\rm SH, \bullet}, \frac{(m_1+m_2)\sigma}{16\sqrt{\pi}n_b\bar {m}^2_b G \tau_{\rm SG} \ln \Lambda}\right\}
\end{align}

\noindent
For our fiducial parameters, 
$a_{\rm widest}=24.9 \ \rm{AU}$ for the segregated case and $200.53 \ \rm{AU}$ for the non-segregated case.
In principle, binaries could soften and be disrupted via encounters during the gas-replenishment episode, however, the GDF hardening described in the following is more efficient at this stage. Therefore binary evaporation due to encounters sets the stage, and determines the SMA of the widest binaries at the beginning of the gas-enrichment stage, but can be neglected during the the time binaries are embedded in gas.

\subsection{Gas dynamical friction}
In gas-rich environments, such as the 2P gas environment of multiple-population GCs/YMCs (and AGN disks), GDF can play a major role in hardening.
The evolution of binaries in gaseous media has been studied over a wide range of astrophysical scales from asteroids to MBHs (as discussed in the introduction).

The effect of gas was suggested to be modeled mainly via several approaches. 
One suggestion is the accretion of gas onto a binary forms a circumbinary minidisk, due to accretion to the Hill sphere. In such disks, torques similar to the ones described type I/II migration of planets in protoplanetary disks could lead to the shrinkage of the binary SMA
(e.g. \citealp{art+91,McKernan2012,Stone2017,Tagawa2020}). Such migration leads to very efficient mergers, far more efficient than the case of interaction dominated by GDF, as we discuss below. 
However, these issues are still debated, and some hydrodynamical simulations show that such torques might lead to outward migration (e.g. \citealp{Moody2019,duf+20,mun+20}), while other hydrodynamical studies indicate that in thin disks one should have inward migration \citep{duf+20,Tiede2020}. We do note that most studies consider initially circular orbits, and generally follow circular orbits, while eccentric orbits could evolve differently, with their orbital eccentricity possibly excited into very high eccentricities, as we discuss below in the context of modeling the evolution through GDF.

Therefore, the approach on which we focus here, considers the effects of GDF \citep{Ostriker1999}. When an object has a non-zero velocity relative to the background gas, the interaction with the gas reduces the relative velocity and therefore hardens binaries (e.g. \citealp{Escala2004,BaruteauCuadraLin2011}). 
The binary hardening induced by GDF for the circular case, with binary components with the same mass $m_1=m_2=m$ is given by \citep{GrishinPerets2016},    

\begin{align}\label{eq: gas dadt}
\frac{da_{\rm bin}}{dt}\bigg|_{\rm GDF}&=- \frac{8\pi G^{3/2}a_{\rm bin}^{3/2}}{\sqrt{m_1+m_2}} \rho_g(t) \frac{m}{v_{\rm rel}^2}f\left(\frac{v_{\rm rel}}{c_s}\right); \\
f(x)&=\begin{cases}
\frac{1}{2}\log \frac{1+x}{1-x}-x, \ 0<x<1, \\
\frac{1}{2}\log\left(x^2-1\right)+\log \Lambda_{g}, \ x>1
\end{cases}\label{eq:f Ostriker}
\end{align}

\noindent 
where $f$ is a dimensionless function derived in \cite{Ostriker1999}, $v_{\rm rel}$ is the velocity of the binary relative to the gas, taken as the Keplerian velocity of the binary, i.e. $v_K=\sqrt{G(m_1+m_2)/a_{\rm bin}}$, which dominates the relative velocity throughout most of the evolution.\\
Under this assumption, eq. \ref{eq: gas dadt} could be written as 
\begin{align}
\frac{da_{\rm bin}}{dt}\bigg|_{\rm GDF}&=- 8\pi\sqrt{\frac{Ga_{\rm bin}^5}{2m}}\rho_g(t) f\left(\frac{v_{K}}{c_s}\right)
\end{align}


For massive binaries, the effect of stellar hardening will be weaker than the effect on less massive stars, as can be seen directly from eq. \ref{eq:stellar hardening}. In contrast, the effect of gas hardening increases with mass (eq. \ref{eq: gas dadt}). Comparison of the two shows that hardening is dominated by gas hardening rather than stellar hardening.
Moreover, although the effect of GDF decreases as the binary hardens, it decays more slowly than the three-body hardening, as could be seen from the scaling $\dot a_{\rm hard, \star}\propto a^2$ and $\dot a_{\rm GDF}\propto a^{3/2}$, and therefore GDF dominates the evolution over stellar-hardening throughout the evolution. After gas depletion, three-body hardening becomes the dominant dynamical process for wide binaries, while for sufficiently small separations, the evolution is GWs-dominated.

\subsection{Gravitational-wave inspiral}
For stellar mass objects GW inspiral becomes important only at very small separations, and can be neglected in regard to main-sequence (or evolved) stellar binaries that merge before GW emission becomes important. However, GW inspiral plays a key-role in the evolution of binaries composed of compact objects.

For a circular binary in the quadruple approximation, the GWs inspiral rate is given by \citep{Peters1964},

\begin{align}
\frac{da}{dt}\bigg|_{\rm GW}=-\frac{64 G^3 m_1m_2(m_1+m_2)}{5c^5a^3}
\end{align}

\noindent
where $G$ is the gravitational constant and $c$ is the speed of light.

Without gas dissipation, the maximal SMA for GW merger within a Hubble time is given by 

\small
\begin{align}
\label{eq:gw_inspiral}
a_{max,GW} = \left(\frac{64\tau_{\rm Hubble} G^3 m_1 m_2(m_1+m_2)}{5c^5}\right)^{1/4}\approx
\nonumber \\
\approx
0.07 \ \rm{AU}\left(\frac{m}{10 \ M_\odot}\right)^{3/4}
\end{align}

\normalsize
A compact binary that is driven by GDF to separations below $a_{max,GW}$ would eventually inspiral and merge, even if it survived the gas-replenishment stage, and would produce a GW-source.

\section{Results}\label{sec: results}

Accounting for the effects of the various processes discussed above, we can follow the evolution of binaries in clusters during the gas epoch and assess its outcomes. Overall we find that under plausible conditions all black-hole binaries initially existing in the cluster inner regions that become embedded in gas during the gas-replenishment phase could be driven to short separations and merge within a Hubble time.

These results suggest that gas-catalyzed GW-mergers in GCs and YMCs, not considered at all in current modeling of GCs,  could serve as an important channel for the production of GW-sources, and plays a key role in the evolution of binaries in such clusters.

Both the GDF and GW-inspiral timescales for lower mass compact objects such as neutron stars (NSs) and white dwarfs (WDs), are longer (as can be seen in eq. \ref{eq:gw_inspiral}), but they are also expected 
to modify their semimajor axis distribution.

Here we focus on mergers of BHs, and postpone a detailed discussion of NS and WD mergers for a follow-up paper, but we should already note that potential WD mergers 
could give rise to the production of explosive events such as type Ia supernovae from mergers of massive white-dwarfs \citep[see also][]{Perets2022}, and could produce GW sources observable by planned GW-detection space missions. NS mergers could produce short gamma-ray bursts and aLIGO GW sources. Combined BH-NS or BH-WD binaries with their high mass but lower mass-ratio could be driven to mergers at intermediate timescales between highest and lowest timescales considered here giving rise to WD/NS disruptions by the BH possibly producing rapid faint SNe \citep[e.g.][and references therein]{zen+19,zen+20,bob+22} or short-GRBs accompanied by a potential GW aLIGO-source. The dynamics of binaries with non-equal masses could be however more complicated and is not explored here.

In the following we discuss our results in detail.

\subsection{Gas-assisted GW-mergers}\label{subsec: separation evolution}

\begin{figure}
    \includegraphics[width=1.\linewidth]{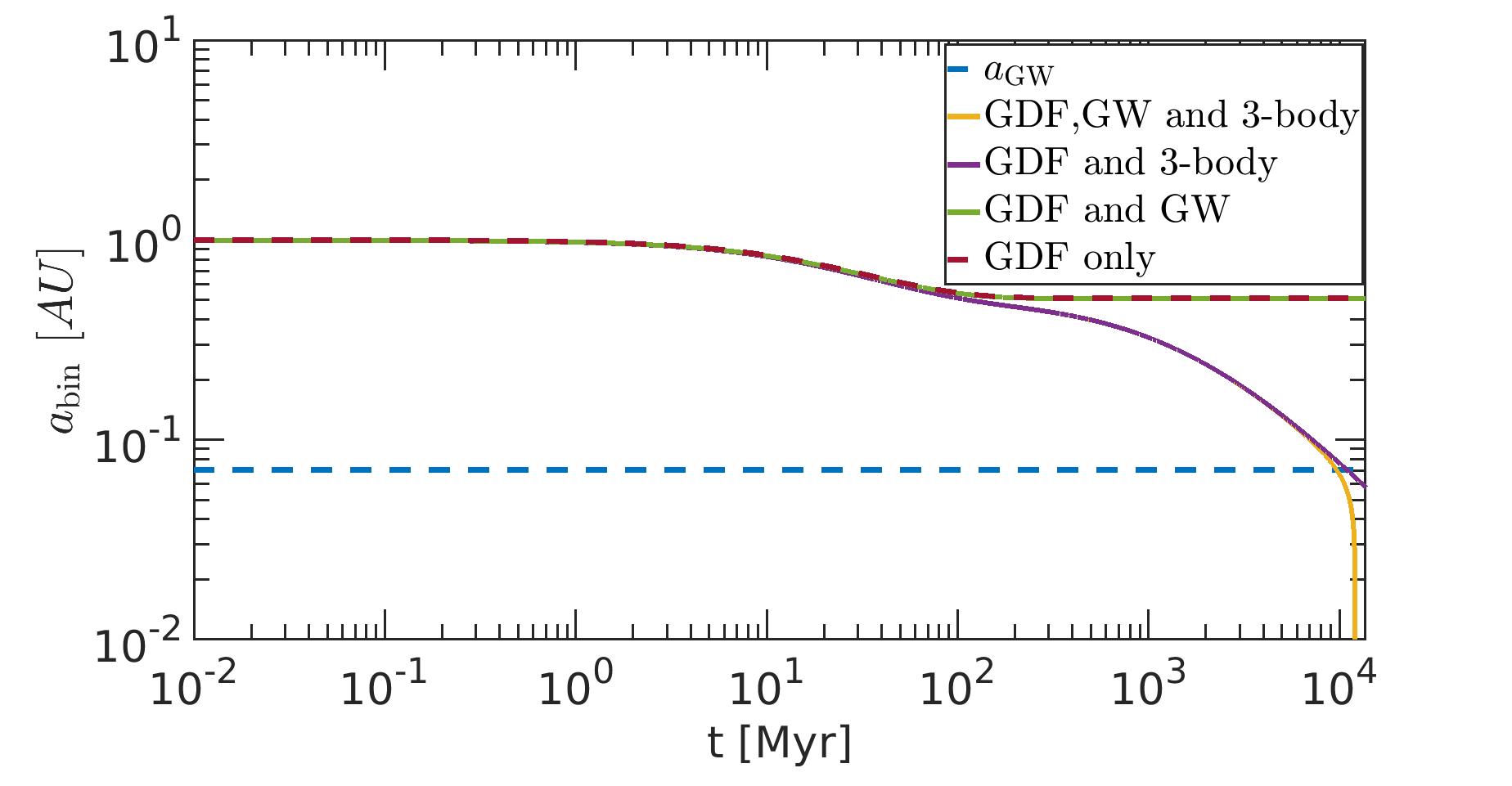}
\caption{The effects of gas hardening, GWs and three-body hardening. The blue dashed line represents the maximal SMA in which GW emission catalyzes a binary merger within a Hubble time.
We consider the evolution of a binary with masses $m_1=m_2=10 \ M_\odot$ and initial separation of 
$a_0=1 \ \rm AU$.
We consider an exponential decaying background gas density $\rho_g = \rho_{g,0}\exp(-t/\tau_{\rm gas})$ with $\rho_{g,0}=1.74\times 10^6 \ M_\odot \rm{pc}^{-3}$ and $\tau_{\rm gas}=50 \ \rm Myr$.
} 
\label{fig:all_together}
\end{figure}

\begin{figure}
    \includegraphics[width=1.\linewidth]{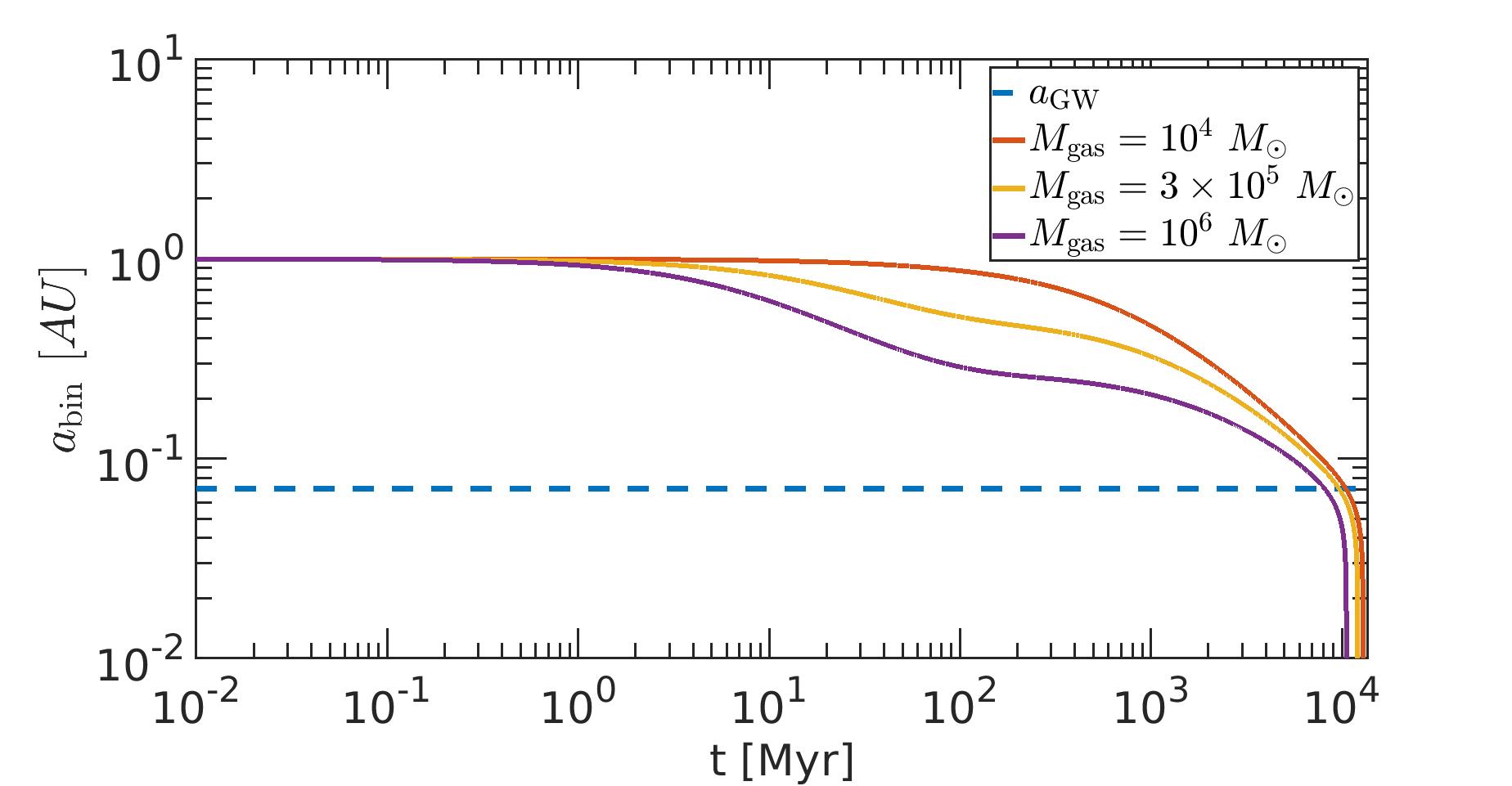}
\caption{
The combined effect of gas hardening, three-body hardening and GWs on a binary, for different background gas masses (and corresponding gas densities). 
The blue dashed line represents the maximal SMA in which GW emission catalyzes a binary merger within a Hubble time.
The solid lines corresponds to the evolution of the SMA, starting from an initial separation of $a_0=1 \ \rm{AU}$, and given different background densities, with an exponential decaying gas density $\rho_g = \rho_{g,0}\exp(-t/\tau_{\rm gas})$ with $\tau_{\rm gas}=50 \ \rm Myr$ (that corresponds to $M_{\rm gas,0}=3\times 10^5 \ M_\odot$). The velocity dispersions are calculated given the total mass of the gas and stars. } 
\label{fig:different densities}
\end{figure}

\begin{figure}
    \includegraphics[width=1.\linewidth]{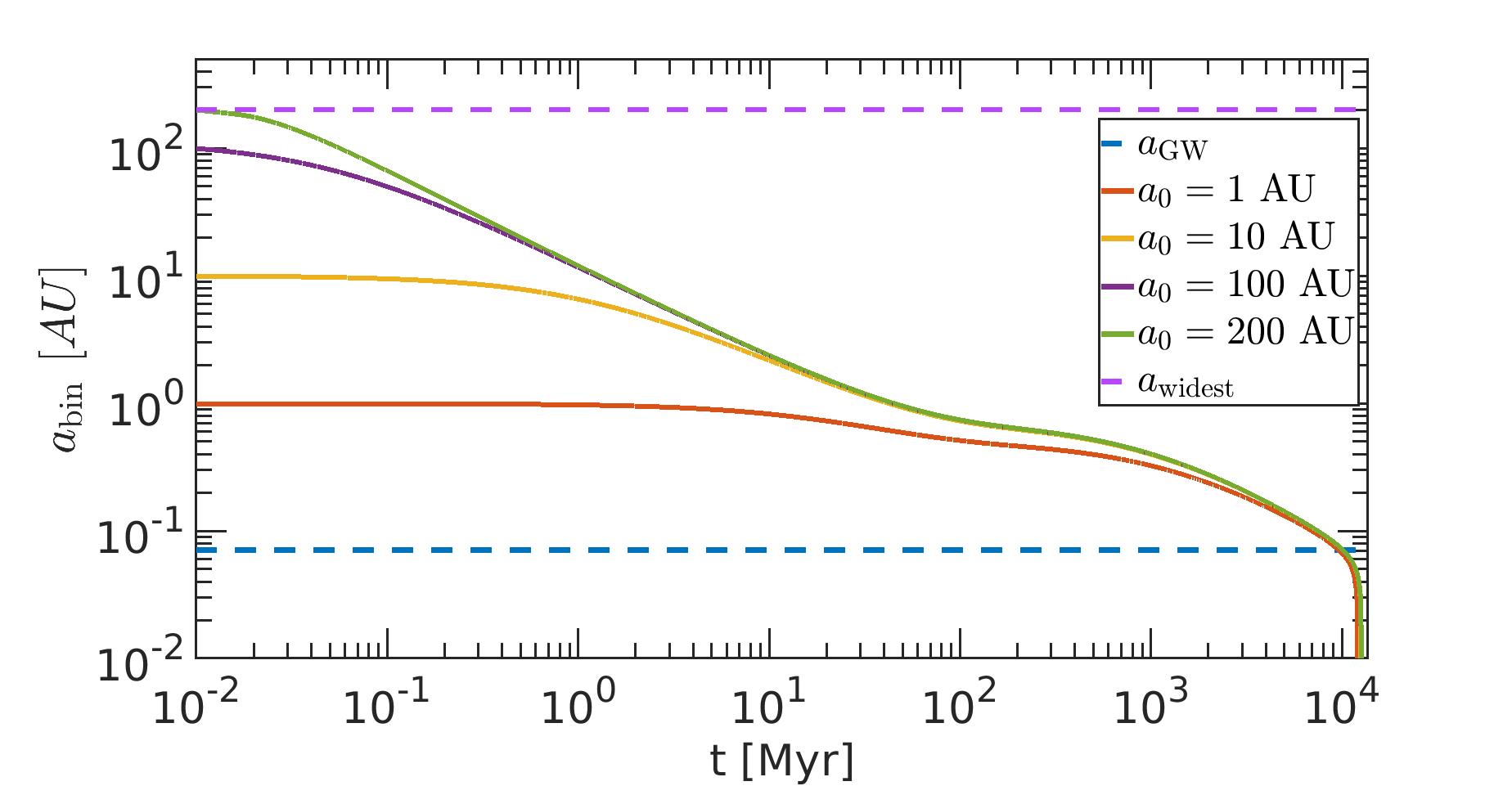}
\caption{
The combined effect of gas hardening, three-body hardening and GWs on a binary, for different initial separations. 
The blue dashed line represents the maximal SMA in which GW emission catalyzes a binary merger within a Hubble time. The purple dashed line corresponds to the widest binary allowed by evaporation considerations.
The solid lines corresponds to the evolution of the SMA, starting from an initial separations of $a_0=1, 10,100,200 \ \rm{AU}$, and given an exponential decaying gas density $\rho_g = \rho_{g,0}\exp(-t/\tau_{\rm gas})$ with $\tau_{\rm gas}=50 \ \rm Myr$.} 
\label{fig:separations}
\end{figure}

In Fig. \ref{fig:all_together} we compare the different hardening processes of binaries in gas-embedded regions. As can be seen, for large separations, the evolution is dominated by the gas hardening, while for smaller separations (at late times after the gas depletion), three-body hardening and finally GWs dominate the evolution.
The transition between the different regimes is determined by the gas density in the cluster, as well as stellar density. Unless stated otherwise, we consider for our fiducial model a background of stars with typical masses of $\bar m=0.5 \ M_\odot$.

In Fig. \ref{fig:different densities} we present the evolution of binaries with an initial separation of $a_0=1 \rm AU$, due to GDF, for different ambient gas-densities. 
The gas hardening mechanism is generally very effective and leads to binary migration to small separations within short timescales, given a sufficiently dense gaseous environment.
As we discuss below, such gas-assisted evolution would then give rise to high rates of GW-mergers of BH binaries, comparable with the BH merger rates inferred from aLIGO-VIRGO-KAGRA (LVK) collaboration \citep{Abbott2016,abb+21}.  

It should be noted that the gas could still dominate the evolution even after reaching $a_{\rm GW}$, as long as the gas was not depleted and the timescale for GWs mergers is larger than the GDF induced merger timescale.
 In principle, GDF-dominated evolution might even be identified in the GW inspiral (in future space missions) before the merger, under appropriate conditions, if GDF still dominates the evolution in LISA frequencies.

We find that circular binaries shrink and reach final small separations, dictated by the initial conditions, which are not sufficiently small as to allow for GW emission alone to drive the binaries to merger even after a Hubble time. Nevertheless, at such short period, these very-hard binaries are more likely to merge due to dynamical encounters on the long-term compared with the primordial population of binary-BHs, and should be appropriately accounted for in simulations of GC stellar populations.

In Fig. \ref{fig:separations}, we introduce the evolution of binaries with different initial separations under the combined effect of GDF, three-body hardening and GWs. It could be seen that although the merger timescales of wider binaries are slightly larger, all the binaries are expected to merge within a Hubble time. Hence, the effect of the presence of gas in the initial stages is robust across all separations and will modify the binary population.
For wide enough binaries, we enter the subsonic range. In order to avoid the discontinuity in eq. \ref{eq:f Ostriker}, we take it as a constant in a small environment around Mach $1$ -- for $\mathcal M<1.01$, we consider $f(\mathcal M)\equiv f(1.01)$, where the widest binary we consider corresponds to $\mathcal M\approx 0.97$.

\begin{figure}
    \includegraphics[width=\linewidth]{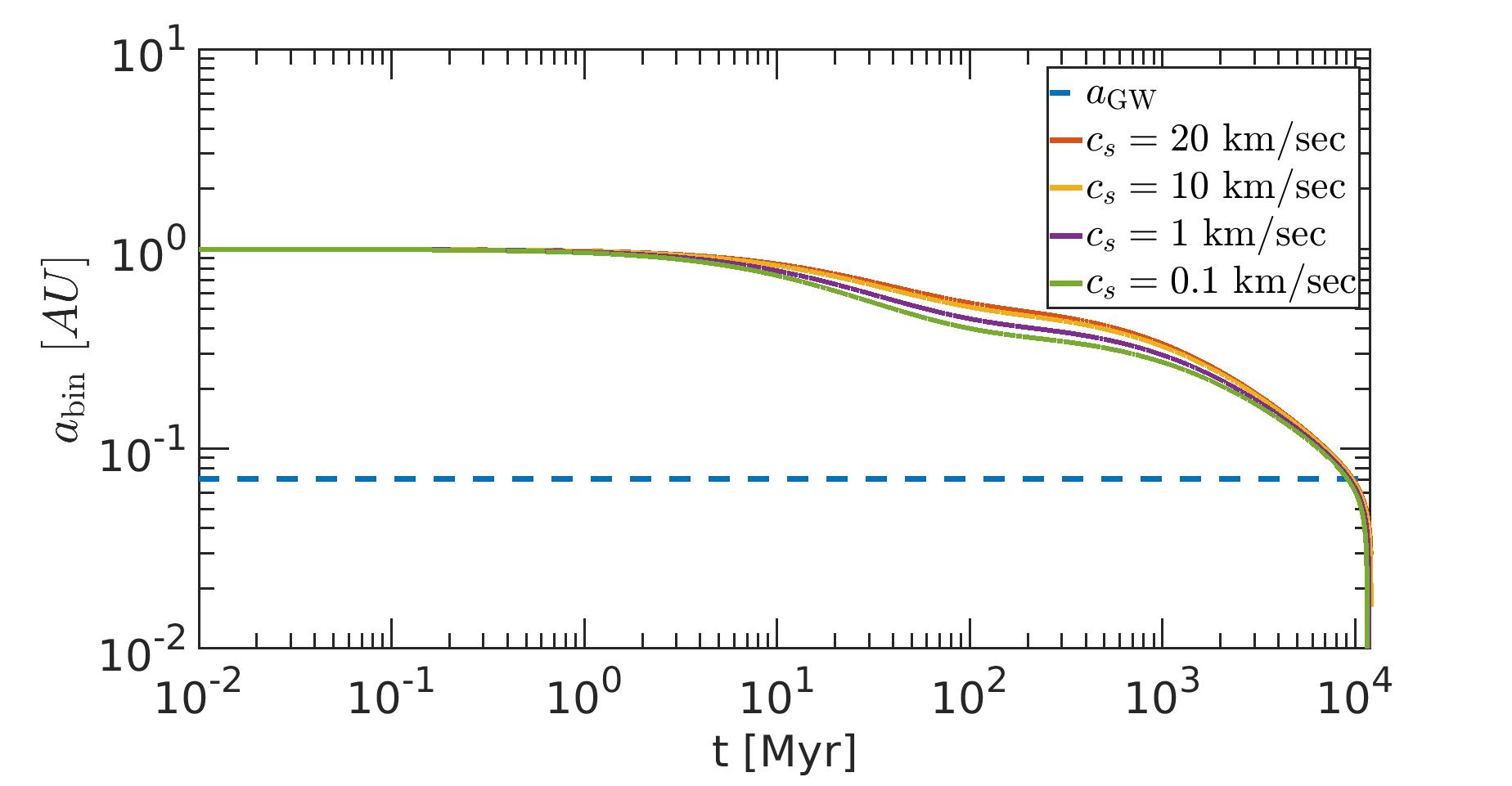}
\caption{
The evolution of the binary separation for different sound speeds. We consider equal mass binaries with initial separation of $a=1 \ \rm{AU}$, masses $m=m_1=m_2=10 \ M_\odot$ and an exponential decaying background density with $\rho_{\rm g,0}=1.74\times 10^6 \ M_\odot \rm{pc^{-3}}$. The blue dashed line corresponds to the maximal separation from which a GWs merger is expected. 
} 
\label{fig:different c_s}
\end{figure}

\begin{figure}
\includegraphics[width=1.\linewidth]{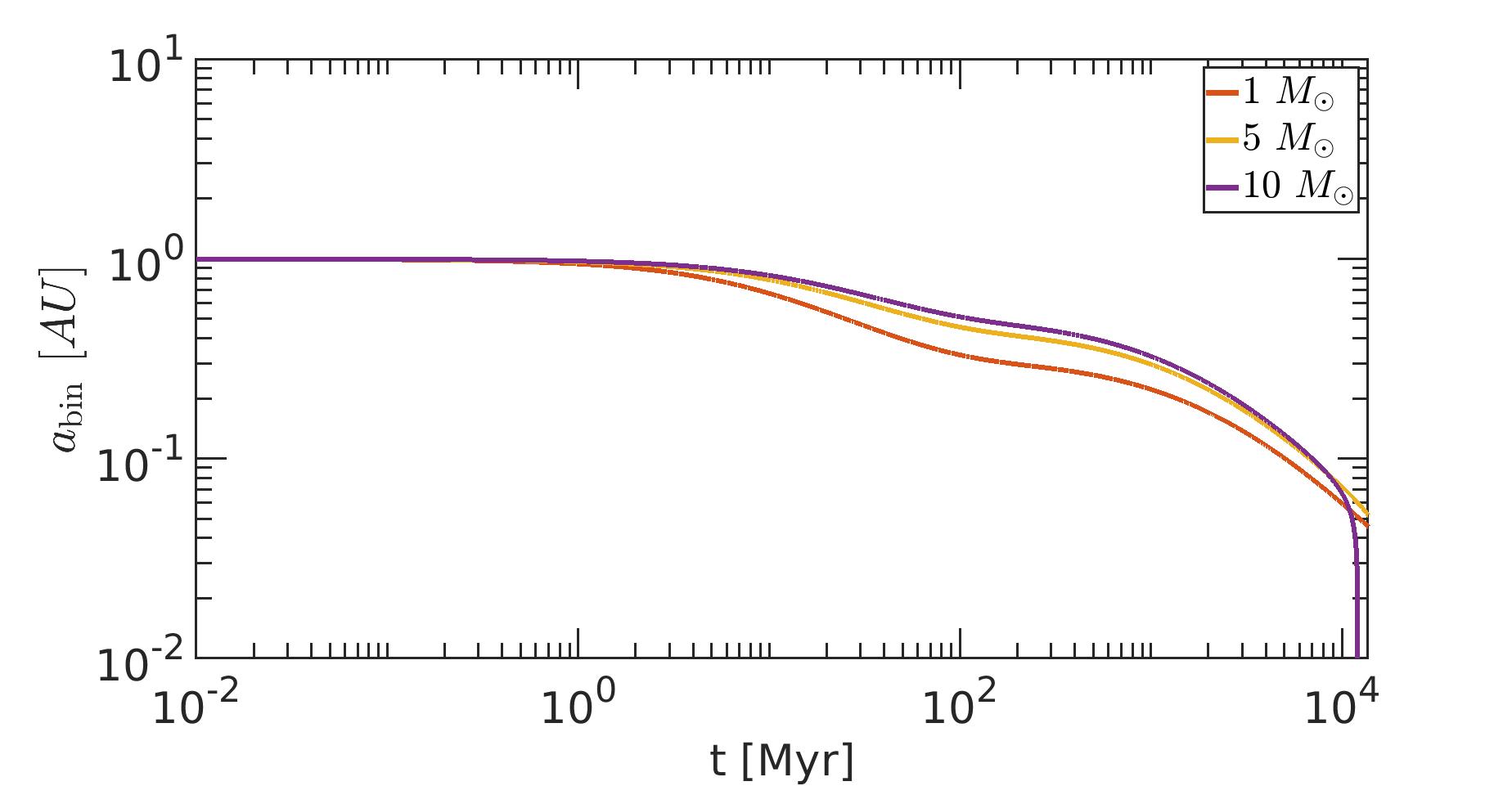}
\caption{
The effect of gas hardening on a binary, as dictated by GDF, for different masses of binaries. 
The different curves correspond to the evolution of the SMA for  different binary masses, starting from an initial separation of $a_0=1 \ \rm{AU}$, given a background density with an exponential decaying gas density $\rho_g = \rho_{g,0}\exp(-t/\tau_{\rm gas})$ with $\tau_{\rm gas}=50 \ \rm Myr$ and $\rho_{g,0}=1.74\times 10^6 M_\odot \rm{pc}^{-3}$.} 
\label{fig:different masses}
\end{figure}

In Fig. \ref{fig:different masses}, we demonstrate the dependence of gas hardening on different binary masses. As can be seen from eq. \ref{eq: gas dadt}, lower-mass binaries harden over longer timescales, due to the dependence on the mass that scales as $\propto \sqrt{m}$, for an equal mass binary with companions $m_1=m_2=m$. The final SMA of the binary also depends on the mass of the binary, such that more massive binaries will attain smaller final SMAs. 

In Fig. \ref{fig:different c_s} we consider different sound speeds, all of them in the supersonic regime. Higher sound speed lead to larger merger timescales, although the results are robust and do not change steeply between the different choices of sound speed in this regime.

  \subsection{Comparison with other gas hardening models}\label{subsec: other models}

Heretofore, we considered gas hardening induced by GDF. However, there are other approaches to model gas hardening. 

In AGN disks, gas hardening is also modeled using processes similar to migration models in protoplanetary disks (as was suggested in the context of AGN disks \citealp{McKernan2012,Stone2017,Tagawa2020}).
Gas is captured in the Hill sphere of a binary and leads to the formation of a circumbinary minidisk. The disk applies a torque on the binary that leads to separation decay similar to migration type I/II in protoplanetary disks, although there were studies that pointed out that this torque could lead to a softening rather than hardening \citep{Moody2019}.
Notwithstanding, we will assume that the formation of a minidisk can take place in GCs and compare the resulted hardening with our GDF model. 
The typical timescale for hardening due to migration torques is given by (e.g. \citealp{McKernan2012}), 
modify to the other type II migration equation

\begin{align}\label{eq:typeII}
\tau_{\rm type II}\sim
46 \ \rm{yr}
\left(\frac{0.01}{\alpha}\right)\left(\frac{0.23}{h/r}\right)^{2}\left(\frac{40 \rm{yr^{-1}}}{\Omega_{\rm bin}(a_{\rm bin})}\right)
\end{align}

\noindent
where $\alpha$ is the Shakura Sunayev parameter, $h/r$ is the aspect ratio and $\Omega_{\rm bin}=\sqrt{G(m_1+m_2)/a_{\rm bin}^3}$ is the angular frequency of the binary. We adopt values of $h/r=0.23$
and $\alpha=0.01$ as a conservative value for the viscosity parameter of the disk. We substitute the $\Omega_{\rm bin}$ that corresponds to a binary with a separation of $1 \ \rm{AU}$. 
Under these assumptions, the migration timescales, which could be used to approximate the hardening timescales, are shorter than the typical migration timescales we derived using the GDF model. These timescales are also shorter then the ones obtained in AGN disks (e.g. \citealp{Stone2017,Tagawa2020}), as expected.
We therefore expect the merger rates we derived to be similar in this case, and even higher for the lowest gas-densities models, where the rates were limited by slower hardening. 
There were more recent studies that suggested modified migration timescales, here taken for an equal mass binary 
\begin{align}
&\tau_{\rm{type II,K}}= \frac{\Sigma_{\rm disk}}{\Sigma_{\rm disk,min}}\tau_{\rm type II}, \\ &\nonumber \Sigma_{\rm disk,min}=\frac{\Sigma_{\rm disk}}{1+0.04 K}, \\ \nonumber &K=\left(\frac{m_1}{m_1+m_2}\right)^2\left(\frac{h}{r}\right)^{-5}\alpha^{-1}
\end{align}
\noindent
These factors lengthen significantly the typical migration timescales, such that for our fiducial model we expect $\tau_{\rm typeII,K}\approx 71515 \ \rm{yr}$. This timescale is still much shorter than the expected timescale calculated via the gas dynamical friction model.

Another approach to modeling gas-induced inspirals is discussed in \cite{AntoniMacLeodRamirezRuiz2019}. 
They simulate Bondi-Hoyle-Lyttelton (BHL) supersonic flows and derive the corresponding energy dissipation, fitted to an analytical theory. While the overall gas hardening timescales could be comparable or shorter for the parameters that are in our major interest, there are significant differences in the scaling. The typical inspiral timescale is given by (eq. $52$ in \cite{AntoniMacLeodRamirezRuiz2019}),

\begin{align}
\tau_{\rm BHL}=61 \ \rm{Myr} &\left(\frac{a_{0}}{AU}\right)^{0.19}\left(\frac{v_{\rm rel}}{\rm{100 \ km \times sec^{-1}}}\right)^{3.38}
\times \\ \nonumber &\times
\left(\frac{20 \ M_\odot}{m_1+m_2}\right)^{1.19}\left(\frac{7.72\times 10^7 \ \rm{cm}^{-3}}{n_{\rm gas}}\right)
\end{align}

\noindent
where $a_0$ is the initial separation of the binary and $n_{\rm{gas}}$ is the number density of the gas, such that $\rho_{\rm gas}=n_{\rm gas} m_p$ where $m_p$ is the proton mass. 

Each model for gas hardening sets a different critical initial separation from which the binary will merge within a Hubble time. 
The timescales dictated both from the type II migration and BHL mechanism are even shorter than the ones expected by our fiducial model. 

Hence, we will conclude that in all the approaches that we considered to model gas hardening, the process is very efficient and leads to a robust rate of mergers, that modifies significantly the binaries' population, while the major difference between them is the time of the merger, dictated by the different gas hardening timescales.

\subsection{Eccentric evolution}
The evolution of binaries in a gaseous medium is significantly different for non-circular binaries. Here we derive and solve the equations for an orbit-averaged eccentric evolution of an initially eccentric binary embedded in gas, but leave a more detailed discussion on the implications for the dynamical 3-body hardening of eccentric binaries to future study. \\
For simplicity, we will assume that the Keplerian velocity of the binary components dominate the relative velocity to the gas, and that the gas velocity is zero relative to the center of mass of the binary. Hence, the relative velocity between the binary and the gas in the center of mass frame is given by
\begin{align}
\textbf{v}_{\rm rel}=\frac{\Omega a}{2\sqrt{1-e^2}}\left[e\sin f \hat r+(1+e\cos f)\hat \varphi\right]
\end{align}
The orbit equations for the GDF for a binary with two equal masses are then given by
\small
\begin{align}
&\frac{{da}}{dt}\bigg|_{\rm GDF}= \frac{2a^{3/2}}{m_{\rm bin}\sqrt{Gm_{\rm bin}(1-e^2)}}\left[F_re\sin f +F_\varphi (1+e\cos f)\right], \\
&\frac{{de}}{dt}\bigg|_{\rm GDF}=\frac{2}{m}\sqrt{\frac{a(1-e^2)}{Gm_{\rm bin}}}\left[F_r\sin f+F_\varphi (\cos f+\cos E)\right]
\end{align}
\normalsize
where $\textbf{F}_{\rm drag}=F_r\hat r+F_\varphi \hat \varphi$, $f$ is the true anomaly and $E$ is the eccentric anomaly.
The orbit-averaged equations are given by
\small
\begin{align}
&\frac{\overline{da}}{dt}\bigg|_{\rm GDF}= 
\frac{4F_0(1-e^2)^2}{\pi m_{\rm bin}\Omega^3a^2}\int_0^{2\pi}
\frac{Idf}{(1+e\cos f)^2\sqrt{1+2e\cos f+e^2}}
,\\
&\frac{\overline{ de}}{dt}\bigg|_{\rm GDF}= \frac{4F_0(1-e^2)^{3}}{\pi m_{\rm bin}\Omega^3a^{3}}\int_0^{2\pi}\frac{I(e+\cos f)df}{(1+e\cos f)^2(1+2e \cos f+e^2)^{3/2}}
\end{align}
\noindent
\normalsize
where $F_0$ is given by $\textbf F_{\rm drag}=F_0I\textbf v_{\rm rel}/v_{\rm rel}^3$.
The orbit-averaged equations for GWs are given by 
\begin{align}
&\frac{\overline{da}}{dt}\bigg|_{\rm GW}=-\frac{64G^3 m_1m_2(m_1+m_2)}{5c^5a^3(1-e^2)^{7/2}}\left(1+\frac{73}{24}e^2+\frac{37}{96}e^4\right),\\
&\frac{\overline{de}}{dt}\bigg|_{\rm GW}=-\frac{304 G^3em_1m_2(m_1+m_2)}{15c^5 a^4 (1-e^2)^{5/2}}\left(1+\frac{121}{304}e^2\right)
\end{align}
\begin{figure}
    \includegraphics[width=1.\linewidth]{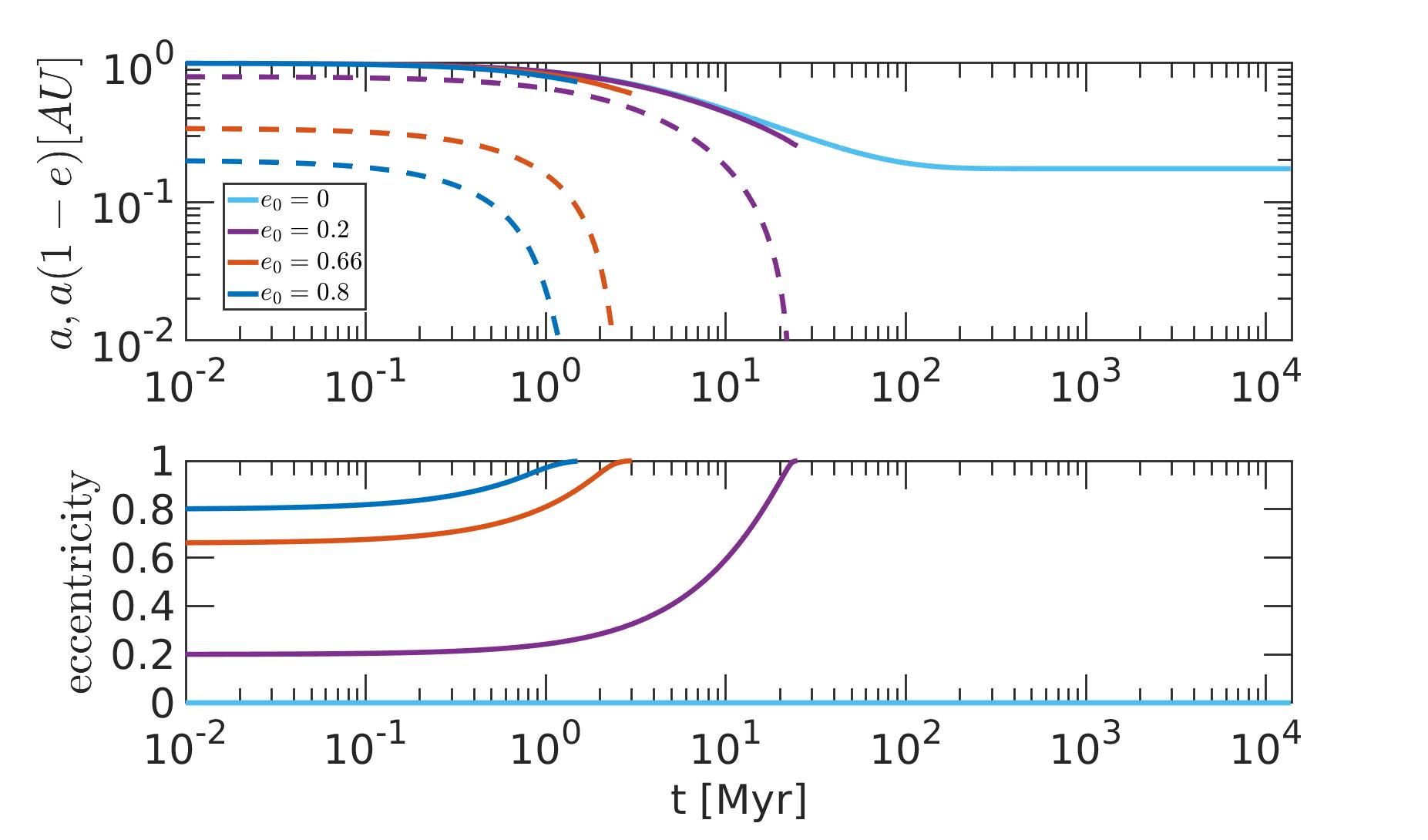}
\caption{The effects of gas hardening on eccentric orbit. 
We consider the evolution of a binary with masses $m_1=m_2=10 \ M_\odot$ and initial separation of 
$a_0=1 \ \rm AU$.
We consider an exponential decaying background gas density $\rho_g = \rho_{g,0}\exp(-t/\tau_{\rm gas})$ with $\rho_{g,0}=1.74\times 10^6 \ M_\odot \rm{pc}^{-3}$ and $\tau_{\rm gas}=50 \ \rm Myr$. The solid lines correspond to semimajor axis evolution and the dashed lines to pericenter evolution. 
} 
\label{fig:eccentric_option1}
\end{figure}
\begin{figure}
    \includegraphics[width=1.\linewidth]{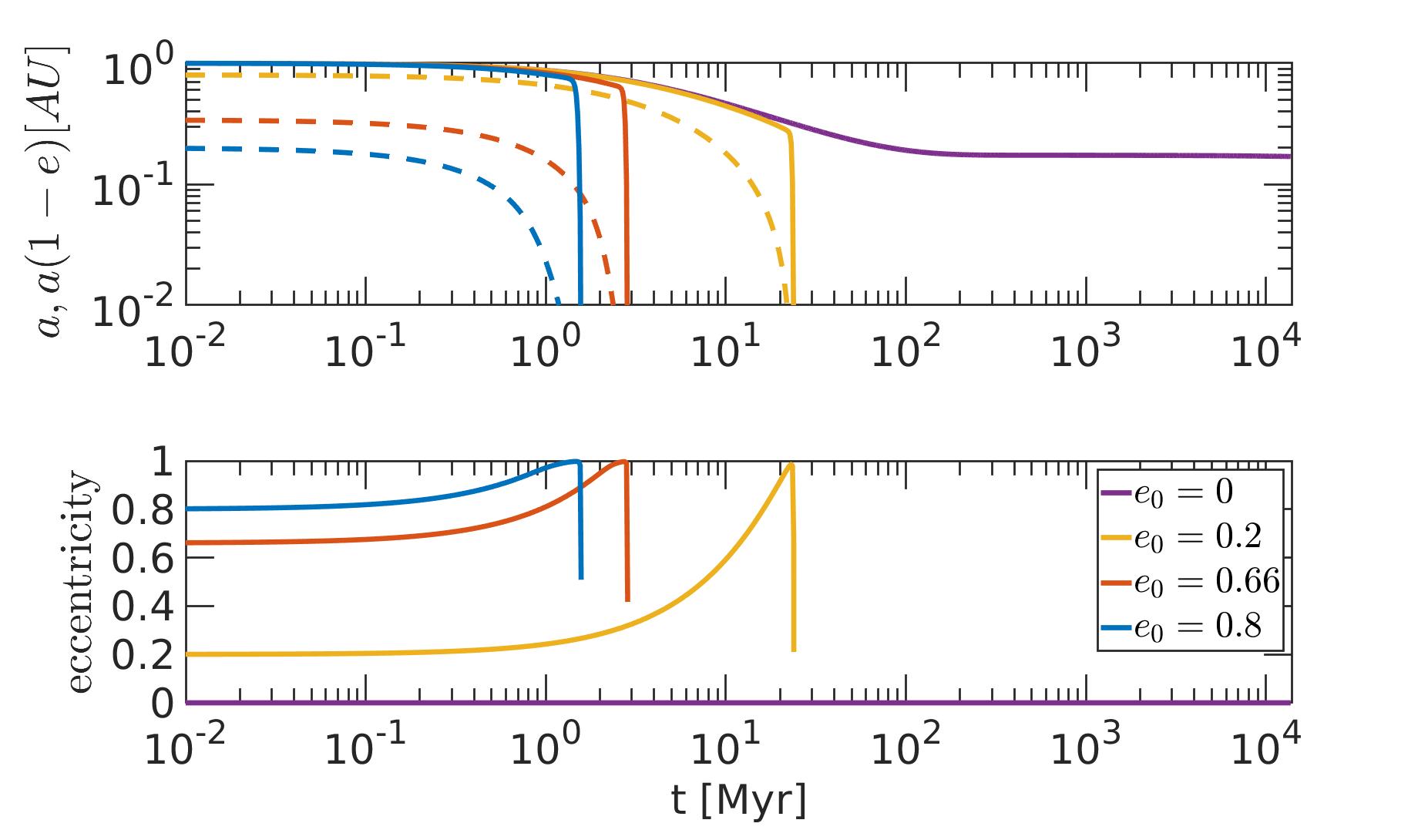}
\caption{The effects of gas hardening and GWs on eccentric orbit. 
We consider the evolution of a binary with masses $m_1=m_2=10 \ M_\odot$ and initial separation of 
$a_0=1 \ \rm AU$.
We consider an exponential decaying background gas density $\rho_g = \rho_{g,0}\exp(-t/\tau_{\rm gas})$ with $\rho_{g,0}=1.74\times 10^6 \ M_\odot \rm{pc}^{-3}$ and $\tau_{\rm gas}=50 \ \rm Myr$. The solid lines correspond to semimajor axis evolution and the dashed lines to pericenter evolution. 
} 
\label{fig:eccentric_+GW}
\end{figure}
In Fig. \ref{fig:eccentric_option1} and Fig. \ref{fig:eccentric_+GW} we introduce the evolution of eccentric binaries.
In Fig. \ref{fig:eccentric_option1}, we present the evolution due only to GDF, and in Fig. \ref{fig:eccentric_+GW}, we also introduce the effect of GW-emission.
As can be seen, the eccentricities become extremely high within short timescales, indicating that the pericenter shrinks significantly. Once the pericenters are sufficiently small, the effect of GWs becomes more significant, and the orbit shrinkage is accompanied by eccentricity damping, and the binaries are driven into approximately circular orbit when entering the VLK GW-bands. 
\\
Such eccentric evolution could play a key role in the evolution of the binary populations, as eccentric binaries merge within potentially far shorter timescales than circular binaries. We note, however, that some studies of a circumbinary gas-disk evolution of binaries, suggest they are only excited to moderate eccentricities \citealp[$\sim0.45$][]{Tiede2020}. Nevertheless, if binaries migration occurs through such processes the overall shrinkage is rapid irrespective of the eccentricity, leading to fast migration timescale (see previous subsec.). 
\\
We further discuss these issues, and in particular the implications for the delay time distribution of GW sources from this channel in subsec. \ref{subsec: mergers rate}. We note that the consideration of eccentric binaries gas-hardening, little studied before should play a similarly important role in binary evolution in AGN disks, possibly in a different manner than in cases where circumbinary-disk evolution is assumed  \citealp{Samsing2020,Tagawaeccentric}.

\subsection{Gravitational-waves merger rate}\label{subsec: mergers rate}

In the following we estimate the GW mergers rate of binary black holes from the gas-catalyzed channel studied here. We will consider old-formed GCs and YMCs separately, given their different formation history. 

In all the models we considered for gas hardening, all the binaries are expected to merge within a Hubble time. However, different gas hardening models suggest different merger timescales.
As discussed above, our GDF models suggest that eccentric binaries merge rapidly, and some of the hydrodynamical studies discussed above suggest that even circular binaries merge during the early gas phase. Since most GCs formed very early, such mergers would not be detected by VLK, given the effectively limited lookback time. However, the younger equivalents of GCs, so called YMCs, continue to form and generally follow the star-formation history in the universe. Hence, mergers in such YMC could occur sufficiently late (and hence closer by) and be detected by VLK and the contribution of YMCs to the total VLK rate will be the dominant one for the eccentric cases (or for all binaries, according to e.g. the circumbinary disk migration models. It should be noted that there is an observational evidence for gas replenishment also in YMCs (e.g. \citealp{Li2016}). If, however, gas densities are lower or the binaries are initially circular/in low eccentricity, the final SMA of the binaries could be larger, leading to longer GW-merger time catalyzed by three-body hardening (driving the delay time distribution to longer timescales), in which case the contribution from old GCs would be the dominant one. 

The rates as a function of the redshift change according to the geometric structure of the 2P stars. Formation of 2P stars in disks is characterized by lower velocity dispersions, that lead to earlier mergers, where for the case of spherical constellation, the higher velocity dispersion leads to later mergers. 

We will start by estimating the number of mergers per cluster, 

\begin{align}\label{eq:f}
N_{\rm merge}\sim  f_{\rm disk}f_{\rm bin,surv}f_{\geq 20 M_\odot}f_{\rm ret}f_{\rm merge}N_\star
\end{align}

\noindent
where 
$f_{\rm disk}$ is the fraction of stars that reside in the disk, 
$f_{\rm bin,surv}$ is the fraction of binaries among massive stars that will survive stellar evolution (i.e. SNe), $f_{\geq 20 M_\odot}$ is the fraction of stars with masses that exceed $20 \ M_{\odot}$, $f_{\rm ret}$ is the retention fraction of BHs in the cluster, $f_{\rm merge}$ is the fraction of binaries that merge among the surviving binaries embedded in the disk and $N_\star$ is the number of stars in the cluster. 

Following our geometrical considerations in subsec. \ref{subsec: disk configuration}, we set $f_{\rm disk}$ in the range $[2\% ,20\%]$. However, even large fractions could be taken into account if there is a significant capture of objects to the disk.

The binarity fraction of massive BHs is $\sim0.7$, although even higher values are quite plausible for the massive star progenitors of black holes (e.g. \citealp{Sana2012}), stellar evolution may reduce this fraction to a typical value of $f_{\rm bin,surv}=0.1$ (e.g. \citealp{AntoininiPerets2012}).
We use a Kroupa mass function for the cluster, such that the fraction of stars with masses larger than $20 \ M_\odot$ is $2\times 10^{-3}$ for non-segregated environment, for segregated ones we take a fraction of $0.01$.
The retention fraction from the cluster is taken to be $10 \%$ (e.g. \citealp{KritosCholis2020} and references therein).
Taking into account the initial survival fraction of wide binaries, we consider $f_{\rm merge}\approx 0.49-0.61$ 
for our fiducial model. The lower value corresponds to massive background stars and the upper limit to low mass background stars ($\bar m=0.5 \ M_\odot$), see a discussion below eq. \ref{eq:widest}.

Following \cite{Rodriduez2016}, we consider logarithmically flat distribution of initial SMA in the range $[10^{-2},10^5] \ \rm{AU}$ where the lower limit is close to the point of stellar contact and the upper one to the Hill radius. It should be noted that although the choice of logarithmically flat is common, there were other choices of distribution considered, based on observational data (see \citealp{AntoininiPerets2012} for further discussion). 

For our fiducial model, $N_\star=10^5$ and $M_{\rm cluster}=10^5 \ M_\odot$.

In order to calculate the GWs merger from old GCs, we follow the calculation of \cite{Rodriduez2016,KritosCholis2020},

\begin{align}
\mathcal R_{\rm old}(z)=\frac{1}{V_c(z)}\int_{z_{\rm min}}^{z}\Gamma_{\rm old}(z')n_{\rm old}(z')\frac{dV_c}{dz'}(1+z')^{-1}dz'
\end{align}

\noindent
where $\Gamma_{\rm old}$ is the rate of mergers in old GCs, $n_{\rm old}$ is the GCs number density, which is taken to be in the range $[0.33,2.57] E^3(z)\ \rm{Mpc}^{-3}$ \citep{PortegiesZwartMcMillan2000,Rodriduez2016,KritosCholis2020}, $dV_c/dz$ is the comoving volume and $(1+z)^{-1}$ accounts for the time dilation. The comoving volume is given by \citep{Hogg1999},

\begin{align}
\frac{dV_c}{dz}&=\frac{4\pi c^3}{H_0^3E(z)}\left(\int_0^z\frac{dz'}{E(z')}\right)^2,
\\
E(z)&=\sqrt{\Omega_M(1+z')^3+\Omega_\Lambda}
\end{align}

\noindent
where $\Omega_K=0, \ \Omega_M=0.3$ and $\Omega_\Lambda=0.7$ \citep{PlanckCollaboration}.

\begin{figure}
    \includegraphics[width=1.\linewidth]{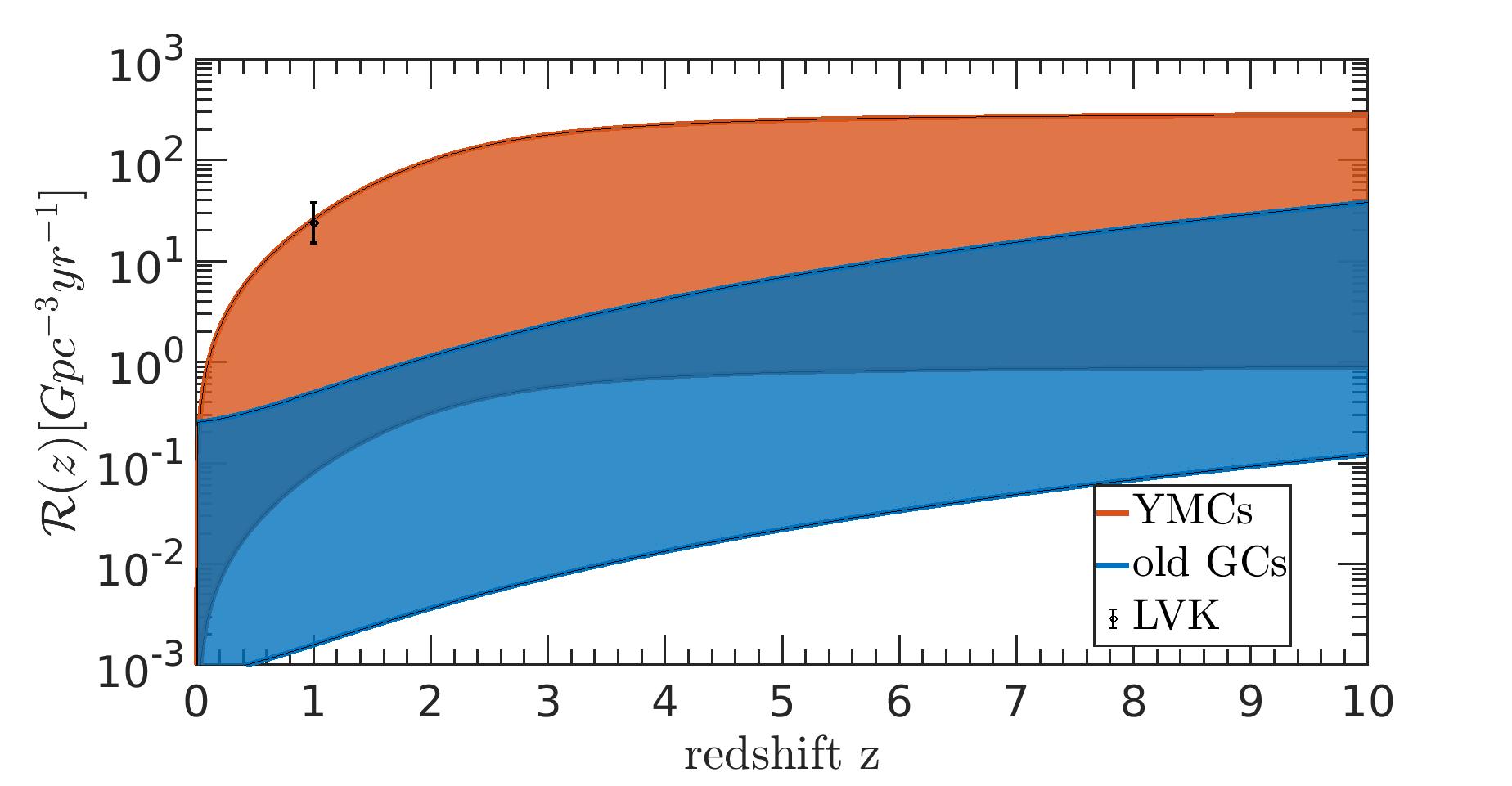}
\caption{The cumulative contribution to GWs rate from YMCs (in red) and old GCs (in blue), from the gas hardening channel, as derived from the GDF. The shaded area relates to the range of parameters. The black line relates to the range of rates inferred by LVK. In the case of circular binaries, the rate will be dominated by old GCs, while for eccentric it will be dominated by YMCs.
} 
\label{fig:rates old}
\end{figure}

As a conservative estimate, we take the mergers rate $\Gamma_{\rm old}$ to be $\Gamma_{\rm old}\sim N_{\rm merge}/\tau_{\rm GC}$ where $\tau_{\rm GC}$ is taken to be $10 \ \rm{Gyr}$. 
In Fig. \ref{fig:rates old} we present the cumulative rate of expected mergers in old GCs (in blue). 
There are two types of contributions to the rate: eccentric binaries, such as these with initial eccentricity of $2/3$ that corresponds to the mean value of a thermal eccentricity distribution, that will merge within short timescales, i.e. with negligible delay time. These practically follow the star formation rate. In this case observed contributions are likely to rise from YMCs. The second case corresponds to low eccentricity/circular binaries, in which there will be a delay time that corresponds to a typical time of $\sim 10^4 \rm{Myr}$. These contributions will be observed in old GCs. 

In this case, the major contribution from our channel to currently observable GW-sources  would not originate from old GCs, but from YMCs.
We define a YMC as a cluster formed later than redshift 2 
and mass $>10^4\ M_\odot$ 
such that we assume for that case that the 2P formation already occurred.
The formation rate of YMCs follows the star formation rate (SFR), which enables us to write the merger rate from YMC as \citep{Banerjee2021V}

\small
\begin{align}
\mathcal R_{\rm young}(z)
=\frac{N_{\rm mrg}}{N_{\rm samp}}\frac{1}{2\Delta t_{\rm obs}}\frac{\int_{M_{\rm cl, low}}^{M_{\rm cl, high}}\Phi_{\rm CLMF}(M)dM}{\int_{M_{\rm GC, low}}^{M_{\rm GC, high}}\Phi_{\rm CLMF}(M)dM}\times \\ \nonumber
\times \frac{\int_{0}^{z}\Psi_{\rm SFR}(z_f)dz_f}{\int_{3}^{6}\Psi_{\rm SFR}(z_f)dz_f}\rho_{\rm GC}
\end{align}
\normalsize
\noindent
$N_{\rm mrg}$ is the number of mergers expected in $N_{\rm samp}$ clusters, $\Delta t_{\rm obs}=0.15 \ \rm{Gyr}$ \citep{Banerjee2021V} is the uncertainty in the cluster formation epoch, $\Phi_{\rm CLMF}\propto M^{-2}$ (e.g. \citealp{PortegiesZwart2010reviewYMC}) is the cluster mass function and we consider $[M_{\rm cl,low},M_{\rm cl,max}]=[10^4, 10^5] \ M_\odot$ as the available mass range for YMCs and $[M_{\rm GC, low}, M_{\rm GC, high}]=[10^5,10^6] \ M_\odot$ as the typical present-day masses for GCs. $\rho_{\rm GC}$ is the observed number density of GCs per unit comoving volume. 
$\Psi_{\rm SFR}(z)$ is the cosmic star formation rate, which is given by \citep{MadauDickinson2014},

\begin{align}
\Psi_{\rm SFR}(z)=0.01\frac{(1+z)^{2.6}}{1+[(1+z)/3.2]^{6.2}}M_\odot \rm{Mpc}^{-3}\rm{yr}^{-1}
\end{align}

\noindent
We consider $N_{\rm mrg}/N_{\rm samp}=N_{\rm merge}$ and spatial densities in the range $[0.33, 2.57] \ \rm{Mpc}^{-3}$, following \citep{Banerjee2021V} and references therein. 
In Fig. \ref{fig:rates old}, we present the cumulative rate of expected mergers in YMCs and GCs. For YMCs, the rate follows the star formation rate (in general, with a small correction due to the delay time -- which is short), and hence peaks in relatively low redshifts. For the eccentric case, the dominant contribution will rise from YMC, while for circular ones the dominant contribution is from GCs.

It should be noted that in general, there could be a non-negligible delay time for the binaries merger. However, for all the parameters we checked for the disk configuration, the merger timescales are extremely short and are negligible in terms of redshifts. 

The total contribution to the GWs merger rate 
from YMCs is in the range $\mathcal R_{\rm young}\approx 
[0.08,25.51] \ \rm{Gpc}^{-3}\rm{yr}^{-1}$, which intersects the expected range of LVK, i.e. $23.9_{-8.6}^{+14.3} \ \rm{Gpc^{-3} \ yr^{-1}}$ \citep{abbott2021},
where the range is bracket by the models with lowest and and highest rates (see Table \ref{table:rates table}).

\begin{table*}[t]
	\begin{tabular}{|c|c||c|c|}
	\hline
		 model &  $\mathcal R_{\rm YMC}(z\leq 1) \ [\rm{Gpc^{-3 }\ yr^{-1}}]$ &  model & $\mathcal R_{\rm YMC}(z\leq 1) \ [\rm{Gpc^{-3 }\ yr^{-1}}]$\\
		 \hline
		 \hline
		 $\rho_{-}c_{s-}n_+$ &  0.32&  $\rho_+c_{s-}n_+$ &  2.55\\
		 $\rho_{-}c_{s-}n_-$ & 0.08 &$\rho_+c_{s-}n_-$ & 0.64\\
		 $\rho_-c_{s+}n_+$ &3.28 &$\rho_+c_{s+}n_+$ & 25.51\\
		 $\rho_-c_{s+}n_-$ & 0.82 &$\rho_+c_{s+}n_-$ & 6.35\\
		 \hline
		\end{tabular}
		\caption{Rates from YMCs for redshifts $z\leq 1$, for different choices of parameters. $\rho_\pm$ correspond to $\rho_{\rm GC}=0.33 E^3(z)\ \rm{Mpc}^{-3}$ and $\rho_{\rm GC}=2.57 E^3(z)\ \rm{Mpc}^{-3}$, $c_{s\pm}$ correspond to $c_{s-}=1 \ \rm{km/sec}$ and $c_{s+}=10 \ \rm{km/sec}$ and $n_\pm$ corresponds to high density of progenitors and low fraction of hard binaries ($n_+$, segregated environment) and low density of progenitors and high fraction of hard binaries ($n_-$, non-segregated environment). These correspond also to different fractions of soft/hard binaries, see subsec. \ref{subsec: soft binaries}. Here we present the rates expected for initially eccentric binaries (e.g. $e_0=0.66$)
		}
		\label{table:rates table}
\end{table*}

In table \ref{table:rates table} we present our calculated rates for different choices of parameters. As expected, higher gas densities lead to larger merger rates and higher sound speeds correspond to thicker disks that host more stars and hence yield more mergers.

\subsection{GW merger properties}
Given the early epoch of gas replenishment, gas-catalyzes mergers operate on primordial binaries in the clusters. The merging components are therefore likely distributed similar to the primordial distribution of binary components. However, even very wide binaries can merge in this channel compared with only relatively close binaries merging in e.g. isolated binary evolution channels for GW mergers. This could give rise to significant differences in the expected masses and mass-ratios of the merger objects.

Interaction with gaseous media could excite binaries to high eccentricities, due to the dependence of the drag force on the relative velocity between the gas and the binary, which changes along the orbit such that the effect is the strongest at the apocenter. Evolution of eccentric binaries hence shorten significantly the expected merger timescales, as larger separations correspond to small pericenters, in which GWs could dominate the evolution. In this case GW-emission would dump the eccentricites and GW-mergers would generally be circular in the VLK bands. However, if the combined  gas-catalyzed and GW-emission binary shrinkage is slower (e.g. for circular-orbits or lower gas-densities), where 3-body encounters dominate the final evolution, and higher eccentricities can be achieved for at least a small fraction of the mergers, similar to the dynamical channels for GW-sources explored in the past.

We should remark in passing  on the possibility of triples. In triples, the outer component migrates faster than the inner binary, potentially leading to an unstable configuration and effective chaotic three body interaction \citep[see e.g. a the reversed case of triples expanding due to mass-loss, leading to similar instability in][]{PeretsKratter2012},
such chaotic encounters could give rise to eccentric mergers. This possibility and its potential contribution will be discussed elsewhere.

\section{Discussion}\label{subsec: discussion}
In the following we discuss 
our results and implications for the evolution of binaries and singles in gas-enriched GCs.

\subsection{Other aspects of binary evolution}

As we showed, the presence of gas modifies the binary population in GCs. It leads to an efficient merger of binaries, together with the formation of binaries via the L2 and L3 mechanism (which was initially used to study the formation of Kuiper-belt binaries \citep{GoldreichSari2002} and recently was applied to calculate the formation rate in AGN disks \citealp{Tagawa2020}).

After the gas dissipation, the initial properties of the binary, as well as the gas, dictate the final separation, to which all the binaries with initial separations larger than the final separations will converge. 

Therefore, gas hardening leaves a significant signature on the binary population and its properties, which sets the ground for further dynamical processes in general and specifically for later dynamical mergers.

In addition to the contribution of the channel to the total rate of GWs, the modification of the properties of binaries (mass, separation etc.) caused by the gas hardening, sets unique initial conditions for the other GWs channels. This will induce an indirect signature of the gas hardening on the expected observed mergers. We introduced analytical results that could in principle be plugged in as initial conditions for the later evolution of GCs, and the dynamical channels for GW production in such environments. The binary abundance changes due to the gas hardening, since a significant fraction of binaries could merge, while others form. Furthermore, additional L2/L3-formed binaries could participate and produce GW sources, beyond the primordial binaries considered here. Nevertheless, since stars might be far more abundant than BHs, L2/L3 processes might mostly produce mixed BH-star binaries and may not contribute to the GW merger rate, but may form other exotic binaries such as X-ray sources etc., and/or produce micro tidal disruption events \citep[][disruption of stars by stellar black holes]{per+16}.  

\subsection{Implications for other gas-rich environments}

The gas-catalyzed dynamics discussed here could take place in any other gas-rich environments, with the proper scaling. While enhanced GW merger rates were discussed in the context of AGN disks \citep[][and references therein]{McKernan2012,Stone2017,Tagawa2020}, they are usually discussed in the context of the evolution of a particular binary or the overall BH-merger rate. However, in those cases too, the whole binary populations, of both compact objects and stars will change their properties.

A very similar process could take place for young binaries embedded in star formation regions \cite{Korntreff2012}. In this case, the effect is limited to a shorter timescale and compact objects might not yet have formed, and are therefore not directly affected (but their progenitor massive stars are).

\subsection{YMCs and very massive clusters}
YMCs are still relatively little studied in the context of the production of GW sources, although their contribution to the total estimated rate of GWs is potentially not negligible \citep{PortegiesZwartMcMillan2000,Banerjee2021V}.  
In these clusters, gas can be present up to smaller redshifts, such that the effect from the channel we suggested for GWs could  potentially be observed. Hence, their overall contribution to the currently observed  merger rates in LVG will be more significant (as can be seen also in Fig. \ref{fig:rates old}). Our rate estimates discussed below, account for both GCs and their younger counterparts, YMCs.

\subsection{Dynamics in gas enriched clusters}
All the dynamical processes that take place in the early stages of GCs evolution might be affected by the presence of gas, e.g. few-body dynamics.

One aspect is that wide binaries that formed during the gas epoch are protected from evaporation by the gas hardening, as they harden within timescales shorter than the typical evaporation/ionization timescales. 

GDF could also enhance mass segregation \citep{Indulekha2013,Leigh2014}. The energy dissipation leads to a change in the velocity dispersion in short timescales, such that massive objects will fall towards the center of the cluster. Moreover, since the more massive objects are prone to merge (as can be seen from eq. \ref{eq: gas dadt}, or visually from Fig. \ref{fig:different masses}), the relaxation will be affected by the modified mass function induced by the gas hardening.

\subsubsection{GW recoils, spins and mass-gap objects}
 
It is possible that gas-accretion onto binaries and not only GDF \citep[e.g][]{Roupas2019} could affect their evolution. In particular, sufficient accretion might align the BH spins and orbits, especially if some circumbinary disk forms around the binaries, in which case the GW-recoil velocity following mergers is likely to be small, and allow a larger fraction of merged, now more massive BHs to be retained in the cluster. This in turn would affect the later dynamics in the clusters, and the resulting mergers in the dynamical formation channels operating in the clusters. This could then potentially give rise to higher fraction of BHs reaching high (even mass-gap) masses following repeated mergers. The spin evolution and accretion, however, require more detailed study, which is beyond the current scope. 

The spin evolution of binaries will be affected by the role played by dynamical encounters, as well as the direction of the gas relative to the binary. In some cases, initially misaligned binaries could be aligned later due to gas accretion, but when dynamical encounters are dominant, the spins won't be aligned.

\subsection{Implications for neutron stars and white dwarfs: accretion \& explosive transients}\label{subsection:accretion}
The focus of the current paper is the merger of BHs and the production of GW sources due to gas interactions in multiple population clusters. However, the evolution of stars and other compact objects such as WDs and NSs could be significantly affected in similar ways. Though some of these aspects are discussed in a companion paper \citep{Perets2022}), we postpone a detailed exploration of these objects to a later stage, and only briefly mention qualitatively some potentially interesting implications.    

A fraction of the gas could be accreted on objects in the cluster. 
Gas accretion changes the velocities of the accretors and the overall mass function of objects in the cluster, such that there is a shift towards higher masses (e.g. \citealp{Leigh2014}), that might affect the dynamical GWs channels in clusters that operate after the gas-replenishment epoch, since we enrich the abundance of massive objects which are likely to be the progenitors of GWs. 
Stars that accrete gas could evolve into compact objects that in turn might produce novae. Enhanced accretion in the early stages of the cluster evolution could potentially modify the novae rates and properties \citep{Maccarone2012} and the production of accretion-induced collapse of WDs into NSs \citep{Perets2022}. 

We should point out that our scenario suggests a robust merger not only of BHs, but also of neutron stars (NSs) and white dwarfs (WDs). These mergers might leave unique signatures. 
Besides their contribution to the production of short GRBs and GW sources, binary NSs mergers are a promising channel to the production of heavy elements via r-process (e.g. \citealp{Freiburghaus1999}), and would affect the chemical evolution of the clusters. 

Thermonuclear explosions of WDs could produce type Ia SNe, whether via single degenerate channel (WD and a non-degenerate companion, \citealp{WhealanIben1973SD}) or double degenerate (two WDs, \citealp{IbenTutkov1984DD}). Both of these channels will be affected by the gas accretion. First, as we mentioned (\citealp{Leigh2014} and references therein), the mass function will change. This is turn might change the characteristics of the SNe and their rate. Furthermore, regardless of the mass variation, a large fraction of the compact object binaries are expected to merge within short timescales, which will also affect the SNe rate. 

Mergers of WDs could yield a remnant merged object with small or absent natal kick and hence constitute another channel for NS formation. 
Accretion could potentially change the retention fraction, and potentially explain the retention problem in the formation of pulsars \citep{Perets2022}. 

\subsection{Constraining the parameters of the cluster}

The amount and origin of gas in GCs during the formation of 2P stars are still uncertain  \citep{Bekki2017}. In this channel, we suggest that the amount of gas dictates a final SMA, such that the separation distribution/GW rate could be used to constrain the gas abundance in the cluster and its lifetime. 

For sufficiently low gas densities (or lower densities following gas depletion), gas hardening is not efficient enough to lead to a merger. 
In this case, the terminal SMA of the binary will exceed $a_{\rm GW}$, such that GWs will not be emitted without a further dissipation process. However, if the gas remains for longer timescales, further hardening will occur. For the whole parameter space we considered, the early stages of the hardening process are very efficient, i.e. wide binaries harden and become hard binaries on short timescales.  

This channel of production of GWs-sources could serve as a tracer to later star formation, as it is coupled to the gas that accompanies this formation. The amount of gas, its decay with time are determined by the star formation history. Since these parameters play a role in gas hardening and hence on the final separation distribution at the end of the gas epoch, they could potentially serve to constrain the 2P gas and star-formation phase, and may help explain some of the differences between 1P and 2P stellar populations.  

For example, we might speculate that the inferred difference between the 1P and 2P binary fractions (e.g. \cite{Lucatello2015}) could be explained by gas-catalyzed hardening and mergers of main-sequence stars residing in the gaseous region. Such 1P binaries which also accrete significant mass of 2P gas would appear and be part of the 2P populations, while outside the gas regions binaries are not affected. In this case some of the 2P binaries preferentially merge compared with 1P stars outside the 2P gas region, leading to an overall smaller binary fractions.

That being said, the many uncertainties and degeneracies involved might be challenging in directly connecting current populations with the early conditions directly.

\subsection{Caveats/future directions}
In the following we discuss potential caveats of our model/scenario. 

$\bullet$ The specific scenario for formation of 2P stars is still unknown/debated, and hence there are large uncertainties in the amount of gas in the cluster, its source during the different stages of evolution. Moreover, some explanations for the different chemical composition of the so-called 2P stars might require lower gas masses than the total mass of 2P stars. In these cases, the phenomena we described might be somewhat suppressed, though, as we have shown even lower gas densities could be highly effective, and will not qualitatively change the results. 

$\bullet$ The expected production rates of GW sources depend on the initial parameters of the clusters we consider, including the gas densities, stellar and binary populations, star formation histories etc. All of these contain many uncertainties, which we did not directly address in this initial study, limited to a small number of models as to provide an overall estimate to bracket the expected GW rates from this channel. Nevertheless, all of our models show that gas-catalyzed mergers in multiple population clusters could produce a significant and even major contribution to the GW-merger rate, and could play a key role in the general evolution of stars and binaries in such clusters.  

$\bullet$ The interaction of gas with binaries is complex and includes many physical  aspects. Here we assumed that the gas density in the cluster, or at least in the region in which the binaries evolve, is spatially constant. 
Most of the gas should be concentrated in the star-forming region, preferentially towards  the inner parts of the cluster. Outer parts of the cluster might be more dilute. Future study could relax the simplified assumption of a constant spatial density and account for a more detailed distribution of gas, stars and binaries.   

$\bullet$ We assumed that the relative velocity between the objects and the gas is dominated by the Keplerian velocity of the binary component become dominant. A more realistic approach, but requiring a detailed Monte-Carlo or N-body simulation could account for the detailed velocity distribution of binaries in the cluster.   

$\bullet$ As we mentioned in subsection \ref{subsection:accretion}, objects embedded in gas could accrete from it and change their mass over time. As a result, their dynamics will change both in the cluster and as binaries \citep{Roupas2019}. Here we considered constant masses throughout the evolution, and neglected the effects of gas accretion. This is a somewhat conservative assumption, in regard to catalysis of mergers, as more massive objects are prown to merge even faster in gas (see eq. \ref{eq: gas dadt} and Fig. \ref{fig:different masses}).

$\bullet$ We considered several choices for the gas depletion, assuming an exponential decay, with a fiducial model of $50 \ \rm{Myr}$ and a  lifetime of $100 \ \rm{Myr}$. However, the formation epochs of stars could set different scenarios, e.g. in which gas is abundant in the cluster for longer timescales of $\sim 100 \ \rm{Myr}$, but only intermittently \citep{Bekki2017}, which will change the picture, or when several wide scale gas replenishmet episode occur over timescale of even many hundreds of Mys or even Gyrs, as might be the case for nuclear clusters. 

$\bullet$ In our analysis we considered for simplicity only equal mass binaries. Though we don't expect a major change in the results, the generalization to binaries with different masses is more complex and requires more detailed population studies, beyond the scope of the current study. 

$\bullet$ It should be noted that there were studies that suggested limited efficiency of gas dynamical friction (e.g. \citealp{Li2020,Toyouchi2020}) than considered here. A more detailed comparison is left out for further studies.
\\
$\bullet$ Although the initial parameters of our disk suggest a thick disk, in later stages the disk will be thinner and finally fragment if it to enable star formation. Hence, for these stages/initial thin disks, the gas hardening epoch should be limited to the regime in which the disk is stable. 
\\
$\bullet$
We restrict ourselves to binaries which are not likely to be disrupted by interactions with other stars. Further disruptions could take place and are encapsulated in $f_{\rm bin,surv}$ (see eq. \ref{eq:f}.

\section{Summary}\label{subsec: summary}
In this paper we discussed the evolution of binaries in gas-enriched environments which likely existed at the early stage multiple-population clusters. We showed the binary interaction with the ambient gas-environment significantly affects their evolution and give rise to major changes in binary population in the cluster and its properties. 

Binaries interaction with gas has been extensively studied over the last few years in the context of AGN disks. Here we show that the environments of multiple population GCs and YMCs similarly give rise to important effects. In particular, focusing on the production of GW sources from binary BH mergers, we find that gas-enriched multiple population clusters could provide a significant and possibly major contribution to the production of GW sources of up to a few tens of Gpc$^{-1}$yr$^{-1}$, comparable with the GW-sources production rate inferred by VLK for the local universe. These might even be higher once formation of new binaries due to gas-assisted capture is considered (to be discussed in a follow-up paper). 

Moreover, we expect catalyzed mergers of other compact objects such as NSs and WDs, and of binary main-sequence and evolved stars to give rise to the enhanced rate of a wide range of merger outcomes, producing a range of transient events such as supernovae, GRBs and the formation of X-ray binaries and stellar mergers, which will be discussed elsewhere.

Furthermore, our findings on the overall evolution of binary populations are relevant for other gas-enriched environments such as AGN disks.

Finaly, our focus here was on binary BH mergers in multiple-population cluster environments, but we point out that the early gas enriched phase of such clusters  (which in practice is relevant to the vast majority of GCs, given that most GCs show multiple populations) significantly affect all the stellar and binary populations, and the overall dynamics inside GCs. Hence the current modeling of the typical initial conditions in GCs and their evolution might need to be fundamentally revised.

\section*{ACKNOWLEDGMENTS}
We like to thank to the anonymous referee for important comments and points that significantly improved the manuscript. We would also like to thank Aleksey Generozov, Johan Samsing ,Jim Fuller, Kyle Kremer, Noam Soker and Evgeni Grishin for fruitful discussions.
MR acknowledges the generous support of Azrieli fellowship. HBP and MR acknowledge support from the from the European Union's Horizon 2020 research and innovation program under grant agreement No 865932-ERC-SNeX. 




\bibliographystyle{aasjournal}






\appendix

\section{Fiducial Parameters}

\begin{table*}[htb]
	\begin{tabular}{c| c| c}
		Symbol & Definition & Fiducial Value\\ 
		\hline
	      	    $\tau_{\rm gas}$ & gas lifetime & 
        $50 \ \rm{Myr}$\\
              $\tau_{\rm SG}$ & formation time of SG & $100 \ \rm{Myr}$\\
	$M_{\star}$ & total mass of stars in cluster & $10^5 \ M_\odot$\\
	$M_{\rm gas}$ & gas mass in the cluster & $3\times 10^5 \ M_\odot$\\
$\rho_{\rm g,disk}$ & initial gas density in disk & $1.74\times 10^6 \ M_\odot \rm{pc^{-3}}$\\
 $h/r$ & scale-height & $0.23$ \\
	$\sigma_{\rm disk}$ & disk velocity dispersion & $ 10 \ \rm{km/sec}$\\
         $\bar m$ & average stellar mass & 
         $0.5 \ M_\odot$
         \\
      $n_\star$ & stellar density & $10^5 \ \rm{pc^{-3}}$
       \\
       $n_{\star,disk}$ & stellar density in disk & $10^5 \ \rm{pc^{-3}}$\\
        $c_s$ & sound speed & $10 \ \rm km/sec$ 
        \\
        $\log \Lambda_g$ & gas Coulomb logarithm & 3.1 
		\end{tabular}
		\label{table:parametres_table_v3}
\end{table*}



\end{document}